\documentclass[]{mosi}

\usepackage{helvet}

\usepackage{amsmath} 
\usepackage{natbib}
\usepackage{graphicx}
\usepackage{subcaption} 

\usepackage[toc,page,header]{appendix}
\usepackage[utf8]{inputenc} 
\usepackage[T1]{fontenc}    
\usepackage{hyperref}       
\usepackage{url}            
\usepackage{booktabs}       
\usepackage{amsfonts}       
\usepackage{nicefrac}       
\usepackage{microtype}      
\usepackage{wrapfig}

\usepackage{amssymb}        
\usepackage{titletoc}
\usepackage{minitoc}

\usepackage{array}
\usepackage{etoolbox}

\definecolor{lightblue}{RGB}{200, 230, 255}  
\definecolor{headerblue}{RGB}{150, 200, 255} 

\usepackage{pgfplots}
\pgfplotsset{compat=1.18}
\usepackage{xcolor}
\usepackage{float}
\usepackage{comment}
\usepackage{multirow}
\usepackage{makecell}
\usepackage{siunitx}
\usepackage{tikz}
\usepackage{pgf-pie}
\usepackage[export]{adjustbox}

\usepackage{ragged2e}
\usepackage{tabularx}
\usepackage{caption}
\usepackage{enumitem}
\usepackage{pifont}
\usepackage[hang,flushmargin]{footmisc}
\usepackage{needspace}

\usepackage{tcolorbox}
\tcbuselibrary{breakable}
\tcbuselibrary{skins}

\usepackage{tabularx}
\usepackage{listings}
\usepackage{threeparttable}
\usepackage{setspace}

\definecolor{MossCyan}{HTML}{82D9FF} 
\definecolor{MossBlue}{HTML}{82B1FF}

\definecolor{ForestGreen}{RGB}{34, 139, 34}
\definecolor{Red}{RGB}{255, 0, 0}

\definecolor{tickG}{rgb}{0.1, 0.588, 0.1}
\definecolor{crossR}{rgb}{0.588, 0.1, 0.1}
\newcommand{\cmark}{\textcolor{green!60!black}{\ding{51}}}
\newcommand{\xmark}{\textcolor{red!70!black}{\ding{55}}}
\definecolor{frenchblue}{rgb}{0.0, 0.45, 0.73}
\definecolor{babyblue}{rgb}{0.54, 0.81, 0.94}
\definecolor{classicrose}{rgb}{0.98, 0.8, 0.91}
\definecolor{beige}{rgb}{0.96, 0.96, 0.86}
\definecolor{forestgreen}{HTML}{2e7d43}

\definecolor{blue1}{HTML}{91BBE6}
\definecolor{blue2}{HTML}{3F90E0}
\definecolor{blue3}{HTML}{316FAD}

\definecolor{color1}{HTML}{FF9999}
\definecolor{color2}{HTML}{FF6666}
\definecolor{color3}{HTML}{FF3333}
\definecolor{color4}{HTML}{E60000}
\definecolor{color5}{HTML}{B30000}
\definecolor{color6}{HTML}{8CD98C}
\definecolor{color7}{HTML}{53c653}
\definecolor{color8}{HTML}{00B050}
\definecolor{color9}{HTML}{2d862d}
\definecolor{color10}{HTML}{206020}
\definecolor{color11}{HTML}{cca300}

\newtcolorbox{promptbox}[2][]{
    colback=white,
    coltext=black,
    arc=3mm,
    boxrule=0.5pt,
    colframe=black!60!white,
    title={#2},
    colbacktitle=black,
    coltitle=white,
    fonttitle=\bfseries,
    top=8pt,
    bottom=8pt,
    left=10pt,
    right=10pt,
    breakable,
    before upper={%
        \linespread{1}\selectfont
        \setlength{\parskip}{1ex plus 0.2ex minus 0.2ex}%
        \setlength{\parindent}{0pt}%
    },
    #1
}

\title{ResearchEnvBench: Benchmarking Agents on Environment Synthesis for Research Code Execution}

\author[2,4\ddagger,5,6,*]{Yubang Wang}
\author[4,5,*]{Chenxi Zhang}
\author[4\ddagger,5,7]{Bowen Chen}
\author[2,5,8]{Zezheng Huai}
\author[2,4\ddagger,5]{Zihao Dai}
\author[1,3,4,5,\dagger]{Xinchi Chen}
\author[4,5]{Yuxin Wang}
\author[1,2,5]{Yining Zheng}
\author[2,5]{Jingjing Gong}
\author[1,2,3,4,5,\dagger]{Xipeng Qiu}

\affiliation[1]{Institute of Trustworthy Embodied AI, Fudan University}
\affiliation[2]{Shanghai Innovation Institution}
\affiliation[3]{Shanghai Key Laboratory of Multimodal Embodied AI}
\affiliation[4]{College of Computer Science and Artificial Intelligence, Fudan University}
\affiliation[5]{OpenMOSS}
\affiliation[6]{Wuhan University}
\affiliation[7]{Nanjing University}
\affiliation[8]{Jilin University}
\contribution[*]{Equal contribution.}
\contribution[\ddagger]{Visiting student.}
\contribution[\dagger]{Corresponding author.}

\checkdata[Repository]{\url{https://github.com/No-518/ResearchEnvBench}}
\checkdata[Correspondence]{\email{xpqiu@fudan.edu.cn}, \email{xc\_chen@fudan.edu.cn}}

\abstract{
Autonomous agents are increasingly expected to support scientific research, and recent benchmarks report progress in code repair and autonomous experimentation. However, these evaluations typically assume a pre-configured execution environment, which requires resolving complex software dependencies, aligning hardware and framework versions, and configuring distributed execution, yet this capability remains largely unbenchmarked. We introduce ResearchEnvBench, a benchmark for environment synthesis in research code execution. Given a research repository, documentation, and a target execution setting, agents must construct an environment that successfully executes at runtime. Evaluations on diverse research repositories reveal a substantial gap in current SOTA agents, with failures dominated by incomplete dependency resolution and brittle version coupling. ResearchEnvBench provides a realistic testbed for advancing autonomous agents toward reproducible scientific research.
}

\begin{document}
\maketitle


\begin{figure}[t]
    \centering
    \includegraphics[width=\linewidth]{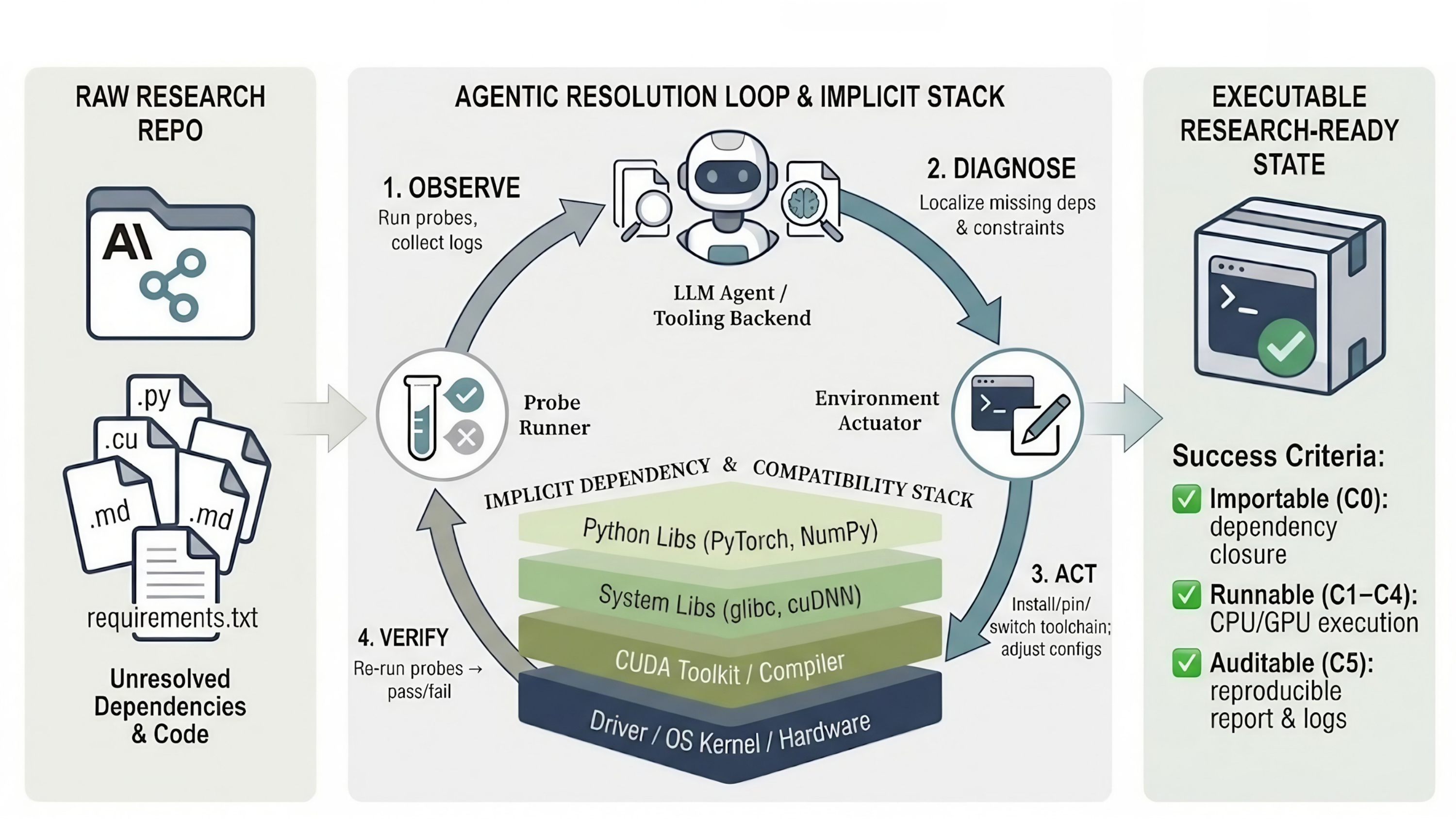}
    \caption{\textbf{Agentic closed-loop resolution of research environments under implicit constraints.} Starting from a raw research repository (left), an LLM-driven agent iteratively observes execution signals, diagnoses missing dependencies and compatibility constraints, acts by installing/pinning packages or switching toolchains, and verifies by re-running probes (center), aiming to reach an executable research-ready state (right). The multi-layer dependency stack highlights that compatibility constraints propagate downward (Python $\rightarrow$ system $\rightarrow$ CUDA $\rightarrow$ driver/hardware), while failures typically surface upward as logs/tracebacks. Evaluation reports importability ($C_0$), runtime CPU/GPU checks ($C_1$--$C_4$), and an auditable, reproducible report and logs ($C_5$).}
    \label{fig:overview}
\end{figure}

\section{Introduction}
Recent work has studied autonomous software engineering and scientific discovery. Agents have demonstrated remarkable proficiency in complex tasks, ranging from repository-level bug fixing~\cite{jimenez2024swebenchlanguagemodelsresolve,yang2024sweagentagentcomputerinterfacesenable} to autonomous machine learning experimentation~\cite{chan2025mlebenchevaluatingmachinelearning,edwards2025rexbenchcodingagentsautonomously}. However, these evaluations predominantly operate within a crucial abstraction: they assume the existence of a functional, pre-configured execution environment. In practical research scenarios, particularly within Deep Learning (DL) and High-Performance Computing (HPC), this assumption rarely holds. Resolving the dependency problem of Python libraries, aligning CUDA drivers with framework versions, and configuring distributed communication primitives, remains a notorious bottleneck (Figure~\ref{fig:overview}). Without the capability to autonomously bridge this pre-execution barrier, an agent's ability to modify code or propose experiments remains theoretically promising but practically unverifiable.

However, benchmarking this environment set up capability remains fundamentally unresolved, particularly for research-grade machine learning codebases. Configuring these environments requires resolving system-level dependencies, such as aligning CUDA drivers with PyTorch versions, compiling custom C++ extensions, and configuring distributed communication primitives. Existing benchmarks fail to capture this runtime complexity. For instance, EnvBench~\cite{eliseeva2025envbenchbenchmarkautomatedenvironment} evaluates setup success primarily through static analysis (checking for missing imports), a metric that cannot detect runtime failures caused by binary incompatibilities or hardware mismatches. Similarly, while Multi-Docker-Eval~\cite{fu2025multidockerevalshovelgoldrush} and Repo2Run~\cite{hu2025repo2runautomatedbuildingexecutable} assess container build success rates, they do not enforce the rigorous runtime verification required to ensure that the constructed environment can actually execute computations on hardware accelerators. This evaluation gap leaves us without a reliable measure of an agent's ability to truly reproduce scientific experiments ``in the wild''.

To bridge this gap, we introduce ResearchEnvBench, a benchmark designed to evaluate the runtime reproducibility of deep learning research environments. Unlike prior datasets that focus on general software packages, we curate a set of 44 high-quality research repositories created after January 1, 2024. These repositories are selected for their complexity, involving heavy hardware dependencies, custom CUDA kernels, and distributed training requirements. To evaluate agent performance, we propose the pyramid of runtime verification, a hierarchical evaluation pipeline that moves beyond simple installation checks. Our protocol strictly enforces a sequence of validation levels: from static dependency integrity, to hardware alignment, and finally to distributed readiness. Furthermore, we introduce metrics to quantify capability hallucination, measuring the discrepancy between an agent's self-reported success and the ground-truth runtime status.

In summary, our contributions are as follows:
\begin{enumerate}
\item \textbf{A Hardened Benchmark for Research Environment Synthesis:} We release \textbf{ResearchEnvBench}, a curated dataset of 44 high-complexity research repositories created after January 1, 2024. Unlike previous benchmarks that filter for general Python projects, our dataset specifically targets repositories with intricate hardware dependencies, custom CUDA kernels, and distributed training requirements, reflecting the challenges of real-world scientific research code.
\item \textbf{The Pyramid of Runtime Verification:} We propose a rigorous evaluation protocol that moves beyond static analysis. We formalize environment setup as a hierarchical capability ladder, requiring agents to pass a sequence of checks ranging from dependency integrity to multi-GPU distributed data parallel (DDP) execution. Additionally, we introduce a capability hallucination metric to quantify the discrepancy between an agent's self-reported success and ground-truth executability.
\item \textbf{Benchmarking SOTA Agents:} We conduct a comprehensive evaluation of four state-of-the-art agent settings on ResearchEnvBench and find a steep drop from hardware alignment to true entrypoint execution (best multi-GPU validation success rate is 37.5\%), alongside large differences in self-report reliability (where conservative \texttt{null} reporting can reduce hallucination false positives compared to optimistic \texttt{ok} claims).
\end{enumerate}

\section{Related Work}

Recent advances have propelled LLM-based agents from function-level code generation~\cite{chen2021evaluatinglargelanguagemodels} to repository-scale software engineering~\cite{jimenez2024swebenchlanguagemodelsresolve,yang2024sweagentagentcomputerinterfacesenable} and autonomous scientific discovery~\cite{yamada2025aiscientistv2workshoplevelautomated,mitchener2025kosmosaiscientistautonomous,ghareeb2025robinmultiagentautomatingscientific}. While agents demonstrate growing proficiency in modifying code logic and iterating on experiments, a critical precursor remains under-evaluated: the ability to autonomously bootstrap the execution environment itself. ResearchEnvBench targets this gap, evaluating whether agents can establish reproducible runtime states for complex research repositories without human intervention.

\subsection{Software Maintenance \& Evolution}

Benchmarks in this category evaluate an agent's ability to modify code logic to fix bugs or add features. SWE-bench~\cite{jimenez2024swebenchlanguagemodelsresolve} established the standard for issue resolution, recently extended by SWE-Smith~\cite{yang2025swesmithscalingdatasoftware} and R2E-Gym~\cite{jain2025r2egymproceduralenvironmentshybrid} which use synthetic data to scale evaluation. Multi-SWE-bench~\cite{zan2025multiswebenchmultilingualbenchmarkissue} and SWE-bench-multilingual~\cite{yang2025swesmithscalingdatasoftware} further expand this to multilingual settings. However, these benchmarks typically operate within pre-built Docker containers where dependencies are already satisfied. They incentivize modifying source code to pass unit tests. In contrast, ResearchEnvBench operates under a no-modification constraint on tracked files, but allows adding auxiliary files, assessing the agent's ability to construct a research-ready environment.

\subsection{Agents for Scientific Discovery}

Automating the machine learning research pipeline is a surging area.
MLE-bench~\cite{chan2025mlebenchevaluatingmachinelearning} evaluates agents on 75 Kaggle competitions, focusing on model performance metrics. RExBench~\cite{edwards2025rexbenchcodingagentsautonomously} and SciCode~\cite{tian2024scicoderesearchcodingbenchmark} assess agents on implementing novel research extensions or solving scientific problems. While these benchmarks evaluate the research capability, they often bypass the ``MLOps'' Challenge by providing simplified environments. ResearchEnvBench complements these works by isolating the environment bootstrap phase. We verify whether agents can handle the specific dependency problem of modern AI research (e.g., FlashAttention compilation, NCCL linking) before any experiment can begin.

\subsection{Automated Environment Setup}

The most relevant recent works focus explicitly on environment configuration. EnvBench~\cite{eliseeva2025envbenchbenchmarkautomatedenvironment} provides a large-scale evaluation for Python and JVM repositories but relies on static analysis (missing imports check) rather than execution. SetupBench~\cite{arora2025setupbenchassessingsoftwareengineering} evaluates system administration tasks like database setup, but focuses on general DevOps rather than Deep Learning stacks. Multi-Docker-Eval~\cite{fu2025multidockerevalshovelgoldrush} and Repo2Run~\cite{hu2025repo2runautomatedbuildingexecutable} assess Docker build success rates across multiple languages. Our work distinguishes itself through its Pyramid of Runtime Verification. Unlike Multi-Docker-Eval, which targets build success, or SetupBench, which targets general services, we explicitly verify GPU-executable runtime correctness. We require agents to prove CUDA Alignment and Distributed Readiness, capabilities that are critical for AI research but absent in general software engineering benchmarks.

\section{ResearchEnvBench}
\label{sec:researchenvbench}
\subsection{Task Formulation}

We formalize ResearchEnvBench as an interactive, multi-turn stage transition problem. The objective is to evaluate an agent's ability to autonomously transform a raw, unconfigured environment into a 
``Research-Readiness'' stage capable of executing complex ML experiments.

\subsubsection{Problem Definition}
Let $\epsilon_0$ denote the initial environment state (a bare-metal Docker container with CUDA drivers but no repository-specific dependencies) and $R$ denote the target research repository. The task is a Markov Decision Process (MDP) where the agent interacts with the environment over T time steps to produce a final state $\epsilon_T$.

\begin{itemize}
    \item \textbf{The Agent Interface (Action Space $A$):} Unlike prior environment-focused benchmarks that restrict agents to static setup scripts such as Installamatic~\cite{milliken2024pipinstallevaluatingllm} or shell-only interaction such as SetupBench~\cite{arora2025setupbenchassessingsoftwareengineering}, ResearchEnvBench provides a compound toolset that combines shell execution with precise file editing, consistent with modern software engineering agents such as SWE-agent~\cite{yang2024sweagentagentcomputerinterfacesenable}. At each step $t$, the agent selects an action $a_t \in A$, which includes:
    \begin{itemize}
        \item \textbf{Shell Interaction:} Executing shell commands to install dependencies and run build steps.
        \item \textbf{File Navigation \& Reading:} Navigating the repository and reading dependency manifests such as \texttt{requirements.txt}.
        \item \textbf{Code Editing:} Editing files to create auxiliary setup scripts but not modifying tracked source code.
    \end{itemize}
    \item \textbf{State Transition:} The environment transitions from state $\epsilon_t$ to $\epsilon_{t+1}$ based on the agent's action $a_t$. The agent receives an observation $o_{t+1}$ consisting of command outputs and file contents, which serves as feedback for the next step.
	    \item \textbf{Goal State:} The goal is to reach a converged state $\epsilon_{final}$ that satisfies the requirement of research implement. Specifically, $\epsilon_{final}$ must be a directly operable Docker image where the repository's training and inference entry points can be executed successfully once the required datasets are available (e.g., downloaded or mounted) without further manual dependency installation or environment configuration.
\end{itemize}

\subsubsection{Verification Protocol} Upon task completion, we freeze the state $\epsilon_{final}$ and inject a suite of hidden verification probes. We evaluate each environment using the \textbf{ResearchEnvBench Pyramid of Validation}, where $C_0$ is an auxiliary static check and $C_1$--$C_4$ are runtime validations:
\begin{enumerate}
    \item \textbf{Static Integrity ($C_0$):} Missing imports reported by \texttt{pyright} (\texttt{reportMissingImports}), reported as missing/total imports.
    \item \textbf{Runtime Integrity ($C_1$):} The environment supports CPU execution of the model's entry point.
    \item \textbf{Hardware Alignment ($C_2$):} The environment correctly maps framework binaries such as PyTorch to the underlying NVIDIA drivers.
    \item \textbf{Single-GPU Computation ($C_3$):} The environment supports actual computation on a single GPU.
    \item \textbf{Distributed Readiness ($C_4$):} (For supported repos) The environment is configured to support multi-GPU distributed data parallel (DDP) execution, a stringent requirement for modern AI research.
    \item \textbf{Hallucination ($C_5$):} Quantifies discrepancies between the agent’s self-report and the ground-truth outcomes from hidden
  probes (paths, versions, and claimed capabilities).
\end{enumerate}

\subsection{Dataset Construction}
\textbf{ResearchEnvBench} targets the long-tail of complex, hardware-sensitive AI research code. We constructed our dataset through a rigorous pipeline that filters for research integrity, hardware awareness, and reproducibility feasibility (Table~\ref{tab:benchmark_comparison}).
\subsubsection{Sourcing and Automated Triage}
We initiated our collection from GitHub, targeting Python repositories created after \textbf{January 1, 2024}. This temporal constraint ensures our benchmark evaluates agents against modern library ecosystems such as PyTorch 2.x, avoiding stale dependency trees common in older datasets.

From an initial pool of candidates, we employed a custom static analysis tool to programmatically filter repositories based on three dimensions of evidence:
\begin{enumerate}
    \item \textbf{Research Integrity Signals:} We filtered for repositories containing explicit research artifacts in their \texttt{README.md}. The triage script searches for research-oriented keywords such as ``arXiv'' to verify the repository is intended for AI scientific contribution rather than general utility.
    \item \textbf{Hardware \& Distributed Semantics:} A unique feature of ResearchEnvBench is the mandatory requirement for hardware acceleration. The triage script scans the file tree for signatures of high-performance computing:
    \begin{itemize}
        \item[$\circ$] \textbf{GPU Dependency:} Presence of GPU-specific keywords such as \texttt{torch.cuda}.
        \item[$\circ$] \textbf{Distributed Training:} Detection of distributed training primitives.
        \item[$\circ$] \textbf{Compilation Complexity:} Identification of CUDA source files, indicating custom operator compilation.
    \end{itemize}
    \item \textbf{Hard Constraints:} To ensure quality, we enforced strict thresholds: at least 100 GitHub stars, no archived projects, and non-forked repositories.
\end{enumerate}

\begin{table*}[t]
    \centering
    \footnotesize
    \setlength{\tabcolsep}{6pt}
    \renewcommand{\arraystretch}{1.15}
    \begin{tabular}{@{}l c c c c l@{}}
        \toprule
        \textbf{Benchmark} & \textbf{Repos} & \textbf{Langs} & \textbf{Domain} & \textbf{HW Aware} & \textbf{Evaluation} \\
        \midrule
        Installamatic~\cite{milliken2024pipinstallevaluatingllm} & 40 & Py & Gen. SWE & \xmark & Build success \\
        EnvBench~\cite{eliseeva2025envbenchbenchmarkautomatedenvironment} & 994 & Py, JVM & Gen. SWE & \xmark & Missing imports \\
        SUPER~\cite{bogin2024superevaluatingagentssetting} & 45 & Py & ML & \xmark & CPU execution \\
        Multi-Docker-Eval~\cite{fu2025multidockerevalshovelgoldrush} & 40 & 9 langs & Gen. SWE & \xmark & Build + tests \\
        SetupBench~\cite{arora2025setupbenchassessingsoftwareengineering} & 54 & 7 langs & DevOps & \xmark & Functional checks \\
        \midrule
        \textbf{ResearchEnvBench (Ours)} & \textbf{44} & \textbf{Py, C++} & \textbf{AI/HPC} & \cmark & \textbf{Runtime pyramid ($C_0$--$C_4$)} \\
        \bottomrule
    \end{tabular}
    \caption[Comparison of ResearchEnvBench with environment setup benchmarks]{Comparison of ResearchEnvBench with state-of-the-art environment setup and execution benchmarks. Prior benchmarks often target general software engineering (Gen.~SWE) or DevOps (development/operations) and evaluate via build success, missing-import checks, unit tests (e.g., \texttt{pytest}), or functional checks; ResearchEnvBench focuses on AI/HPC research and uses a hardware-aware runtime pyramid ($C_0$--$C_4$).}
    \label{tab:benchmark_comparison}
\end{table*}

\subsubsection{Final Dataset Composition}
\begin{figure}[t]
    \centering
    \includegraphics[width=\linewidth]{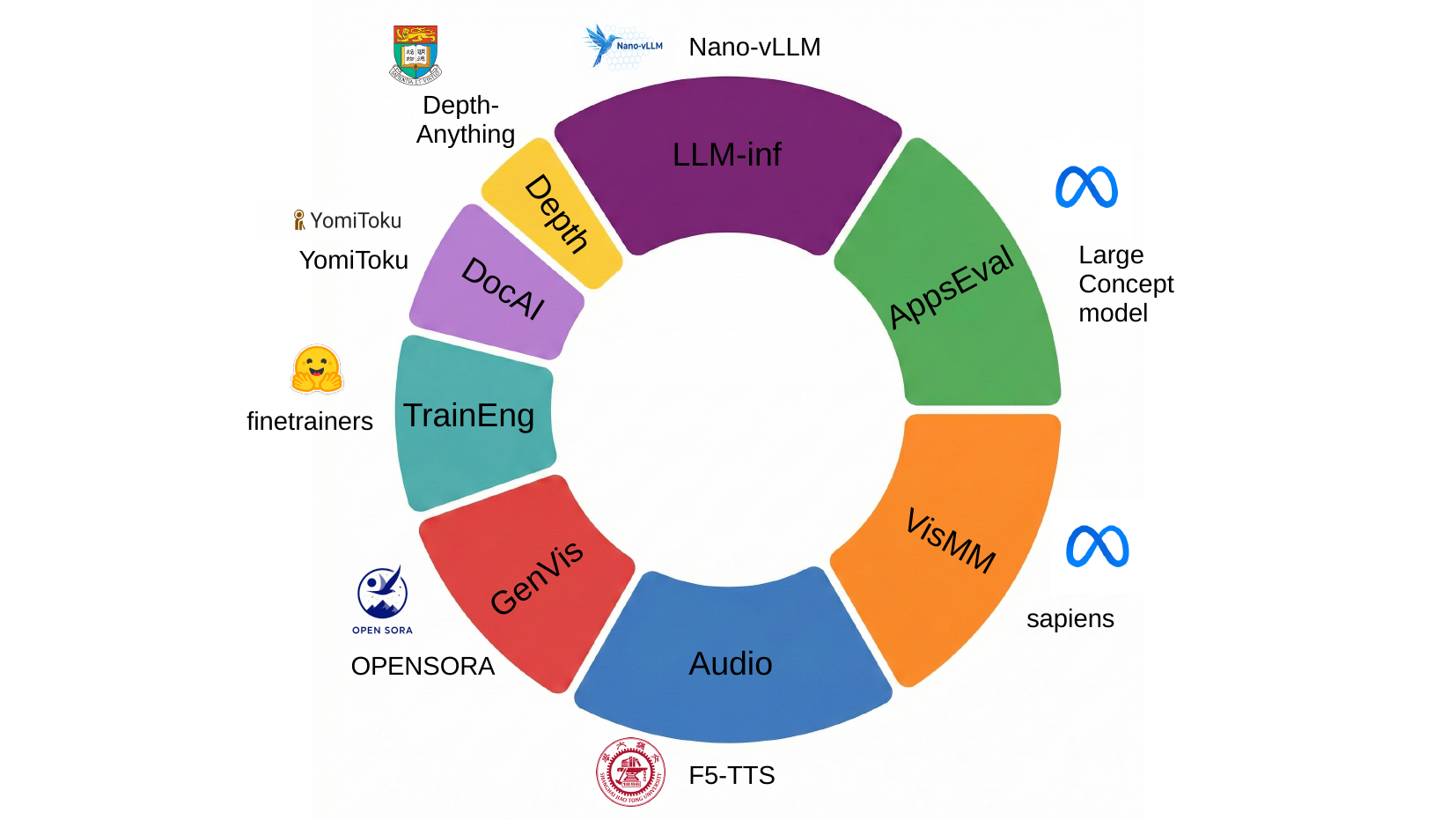}
    \caption{\textbf{Repository composition of ResearchEnvBench.}
    We curate 44 post-2024 ML research repositories spanning eight categories:
    GenVis (generative vision), Depth (depth estimation), Audio (audio \& speech),
    LLM-Inf (LLM inference \& acceleration), TrainEng (training \& engineering frameworks),
    VisMM (vision/multimodal foundations), DocAI (document AI / OCR / translation),
    and AppsEval (applications \& evaluation).
    Slice sizes denote the fraction of repositories per category by count, and each slice shows a representative flagship repository from that category.}
    \label{fig:repo-composition}
\end{figure}

From the candidates, 44 repositories passed all automated checks and manual verification for execution feasibility. While smaller in raw count compared to static analysis benchmarks, this scale is consistent with state-of-the-art execution-based benchmarks such as Multi-Docker-Eval~\cite{fu2025multidockerevalshovelgoldrush} (40 repos) and SetupBench~\cite{arora2025setupbenchassessingsoftwareengineering} (54 repos).

Crucially, ResearchEnvBench prioritizes validation depth over breadth. Nearly all selected repositories in our final set (43/44) support at least single-GPU execution, and a substantial subset additionally supports multi-GPU distributed data parallel (DDP) execution. The final dataset spans eight categories of modern ML research codebases (Figure~\ref{fig:repo-composition}), with proportions computed by repository count. Overall, it covers both model-centric workloads (e.g., generative vision, depth estimation, audio/speech) and system-centric stacks (e.g., LLM inference/acceleration and training infrastructure), which jointly stress dependency closure, CUDA alignment, and distributed readiness.

This diversity ensures agents are tested against a wide spectrum of dependency graphs, ranging from standard PyTorch ecosystems to complex environments requiring the compilation of custom kernels and distributed communication primitives.
\subsection{Evaluation Metrics \& Pipeline}
To rigorously assess the ``Research Readiness'' of the synthesized environments, we design a hierarchical runtime verification pipeline and a novel metric to quantify agent hallucinations.
\subsubsection{The Pyramid of Runtime Verification}
\begin{figure}[t]
    \centering
    \includegraphics[width=\linewidth]{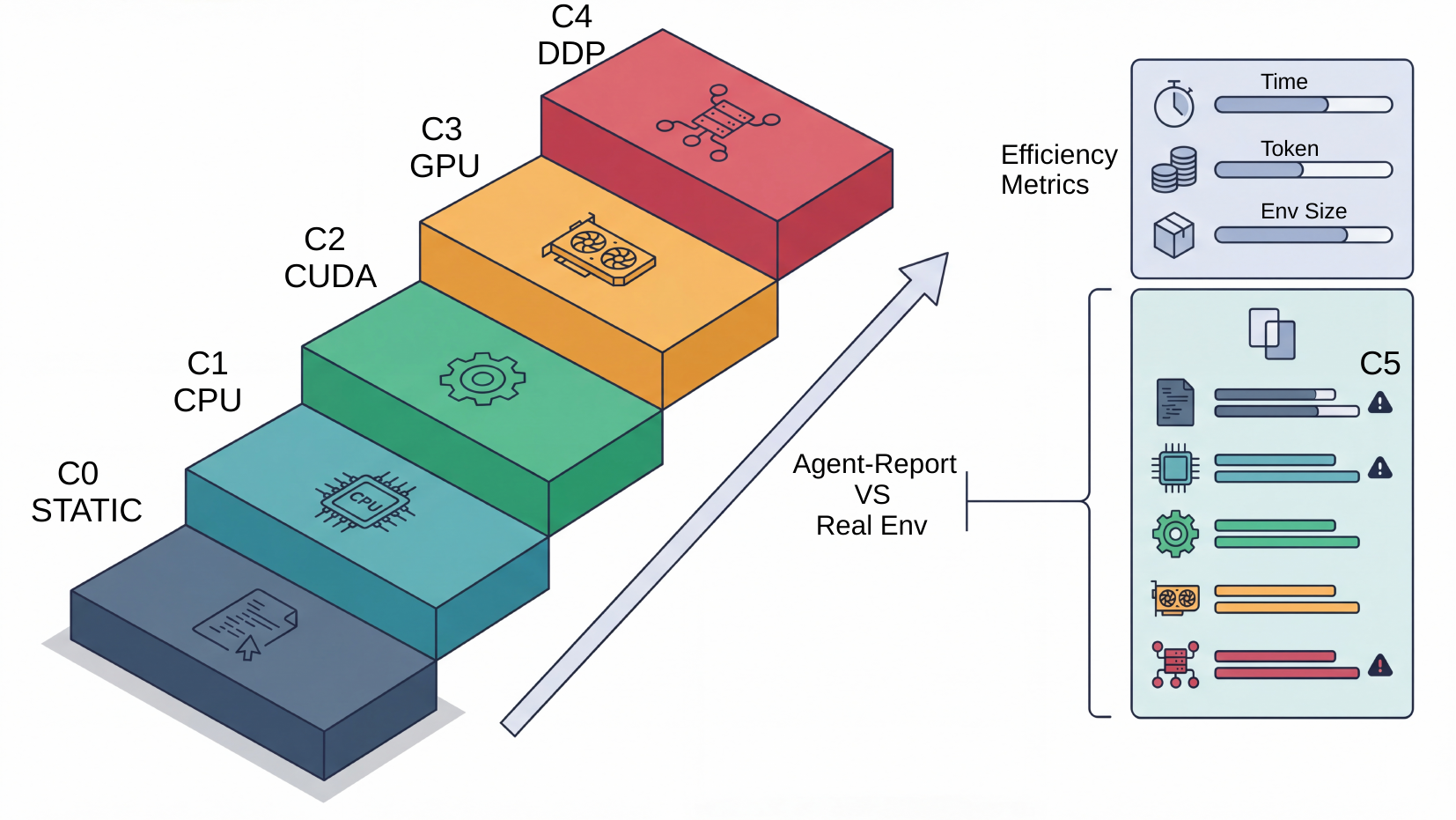}
    \caption{\textbf{The Pyramid of Runtime Verification.}
    We structure environment synthesis into a hierarchy of increasing complexity, ranging from static dependency resolution ($C_0$) to multi-GPU distributed training ($C_4$).
    The right panel illustrates our auxiliary metrics: \textit{Efficiency} (Time, Token, Size) and \textit{Capability Hallucination} ($C_5$), which quantifies the discrepancy between the agent's self-reported success (Report) and the ground-truth execution (Real Result).}
    \label{fig:pyramid}
\end{figure}

Unlike prior works that rely on static build success signals such as Docker exit codes~\cite{fu2025multidockerevalshovelgoldrush}, ResearchEnvBench enforces a strict, multi-stage protocol visualized in Figure~\ref{fig:pyramid}. The execution is orchestrated by a serial protocol (\texttt{run\_all.sh}) that evaluates a hierarchy of environment capabilities. We verify correctness across $C_0$--$C_4$ (increasing runtime strictness) and additionally measure $C_5$ (capability hallucination), which captures false success claims.

\begin{itemize}
		    \item \textbf{$C_0$: Static Dependency Integrity.} Following EnvBench~\cite{eliseeva2025envbenchbenchmarkautomatedenvironment}, we run \texttt{pyright} and count missing-import diagnostics (\texttt{reportMissingImports}), reported as missing imports over total imports. We treat this as an auxiliary proxy for dependency closure; overall readiness is determined by the runtime stages ($C_1$--$C_4$).
    \item \textbf{$C_1$: Minimal CPU Execution.} Following environment synthesis, we require the model's entry point (training or inference) to execute on CPU under the smallest feasible configuration. This validates that the resolved dependency graph is internally consistent, independent of hardware acceleration.
    \item \textbf{$C_2$: Hardware Alignment.} We verify that the installed deep learning framework matches the underlying driver version. The environment must pass a cuda-check probe, ensuring it can access the GPU device.
    \item \textbf{$C_3$: Single-GPU Computation.} The environment must support actual kernel execution on a GPU. This stage catches common version mismatches (e.g., \texttt{torch} compiled with an incompatible CUDA version) that only surface during tensor allocation.
    \item \textbf{$C_4$: Distributed Readiness.} For repositories supporting distributed training, we enforce a multi-GPU check using standard launchers. This verifies the correct configuration of distributed communication primitives like NCCL, representing the highest standard of research reproducibility.
\end{itemize}

\subsubsection{Applicability and Scoring} Not all repositories admit every capability. For repositories without a runnable CPU-only entry point, we treat $C_1$ as skipped; for repositories without multi-GPU distributed support, we treat $C_4$ as skipped. Skipped capabilities are excluded from the denominator when computing aggregate success rates.

\subsubsection{Quantifying Capability Hallucination (\texorpdfstring{$C_5$}{C5})}
Agents often exhibit overconfidence, claiming tasks are complete when underlying configurations are broken. We introduce \textbf{Capability Hallucination} to quantify the discrepancy between the agent's self-report and the observed reality.

The agent is required to generate a structured report summarizing the environment state after synthesis. We classify hallucination into three categories:
\begin{enumerate}
    \item \textbf{Path Hallucination:} The reported python path does not exist or is not executable.
	\item \textbf{Version Hallucination:} The reported library versions like torch\_version differ from the actual versions detected by our hidden probes.
	\item \textbf{Capability Hallucination:} A critical mismatch where the agent claims a capability is enabled but the corresponding runtime verification (e.g., $C_2$ or $C_4$) fails. 
    We record a hallucination whenever the agent claims the check
    after environment setup passes but the probe fails. This metric directly measures how trustworthy an agent's completion claims are.

\end{enumerate}
\subsubsection{Efficiency Metrics}
Beyond correctness, we measure the cost of environment synthesis:
\begin{itemize}
    \item \textbf{Time-to-Ready:} Total wall-clock time from task initiation to agent execution end.
    \item \textbf{Token Cost:} Total number of tokens consumed during the interaction.
    \item \textbf{Environment Size:} Total size of the constructed environment, encouraging lightweight dependency management.
\end{itemize}

\section{Experiment}

We conduct a comprehensive evaluation of state-of-the-art LLM agents on ResearchEnvBench to assess their ability to bootstrap complex ML research environments from scratch.

\subsection{Experimental Setup}
\subsubsection{Evaluation Environment} To ensure reproducibility and prevent environment contamination, each evaluation session is instantiated in a strictly isolated Docker container. The environment is initialized as a clean sandbox based on Ubuntu 22.04, equipped with 2× NVIDIA RTX 4090 GPUs (24GB VRAM each) and CUDA 12.4 drivers (Version 550.163.01). Crucially, the sandbox contains no pre-installed deep learning frameworks (e.g., PyTorch, TensorFlow) or build tools (e.g., CUDA Toolkit), testing the ability of agents to install these from scratch.

\subsubsection{Agent Execution Loop} We utilize a standardized Host Orchestrator to manage the agent lifecycle. For each repository, the orchestrator initializes a clean Docker container and clones the target repository, then provides the agent with a standard interface supporting shell execution, file reading, and file editing. The agent is constrained by a maximum budget of time. Upon the execution completion signal (or upon timeout), the orchestrator immediately revokes access and executes the pyramid verification Pipeline (C0→C5) described in Section~\ref{sec:researchenvbench} via run\_all.sh.

\subsection{Baseline Agents}

We evaluate four agent settings representing distinct capabilities in the current LLM landscape. All agents are prompted with the same system instruction: to act as a ``Senior MLOps Engineer'' focused on reproducibility and verifiability.

\begin{enumerate}
    \item  \textbf{Claude Code Agent (GLM-4.7):} An agent built using Claude Code tooling~\cite{anthropic2026claude_code} with GLM-4.7 as its underlying model. GLM-4.7 is positioned as a flagship model optimized for agentic coding, long-horizon task planning, and tool use; its model card reports strong coding benchmark performance (e.g., 84.9 on LiveCodeBench V6, 73.8\% on SWE-bench Verified, 66.7\% on SWE-bench Multilingual, and 41\% on Terminal Bench 2.0), and claims first-place ranking among open-source and domestic models on Code Arena~\cite{zhipu2025glm47}.
    \item \textbf{Claude Code Agent (Sonnet 4.5):} The same Claude Code agent tooling~\cite{anthropic2026claude_code} with Claude Sonnet 4.5 as its underlying model, evaluated under the identical interface and budgets.
	    \item \textbf{Codex Agent (GPT-5.1-Codex):} An agent powered by GPT-5.1-Codex and provided with the same tool interface as other agents. We treat it as a code-specialized baseline (Codex-style~\cite{chen2021evaluatinglargelanguagemodels}) and evaluate whether extensive code-centric pre-training translates to effective system administration and runtime debugging skills.
    \item \textbf{NexAU Agent (DeepSeek-V3.1-Nex-N1):} An agent built with the NexAU agent framework~\cite{agiteam2025nexn1agenticmodelstrained}, using DeepSeek-V3.1-Nex-N1 as its underlying model. It orchestrates iterative plan-execute-verify loops over build and runtime tooling, enabling the agent to inspect build logs, diagnose compilation failures, and interactively update environment configurations and dependency constraints.
\end{enumerate}

\subsection{Main Results}

Table~\ref{tab:main_results} summarizes missing-import statistics ($C_0$), stage success rates ($C_1$--$C_4$), and self-report hallucinations ($C_5$).

\textbf{Stage-wise executability.} No single agent dominates across the runtime pyramid. Codex attains the highest minimal CPU execution rate ($C_1$, 58.6\%), NexAU and Claude (Sonnet 4.5) tie for the highest CUDA alignment rate ($C_2$, 93.2\%), and Claude variants attain the highest DDP success rate ($C_4$, 37.5\%). Notably, Claude Code (Sonnet 4.5) attains higher $C_2$ than Claude Code (GLM-4.7) (93.2\% vs.\ 90.9\%) but lower $C_1$ and $C_3$ (51.7\% vs.\ 55.2\%; 41.9\% vs.\ 48.8\%). Despite these differences, all agents exhibit a steep drop from CUDA alignment ($C_2$) to true entrypoint execution ($C_3$/$C_4$), reinforcing that ``GPU-visible'' stacks are often not repository-ready.

\textbf{Static integrity is not sufficient.} Although Codex achieves the lowest missing-import ratio under pyright ($C_0$, 23.6\% missing imports), this does not translate into the best GPU readiness: it underperforms on $C_2$ (79.5\%) compared to Claude Code (90.9--93.2\%) and NexAU (93.2\%). This gap indicates that dependency closure alone does not guarantee correct accelerator builds and ABI alignment.

\textbf{Hallucination behavior.} $C_5$ reveals large differences in self-report reliability. Codex produces only 4 total hallucinations, while Claude Code (GLM-4.7), Claude Code (Sonnet 4.5), and NexAU produce 18, 20, and 16, dominated by capability hallucinations (17/18, 20/20, and 14/16). In manual inspection, Codex is more conservative in its report and often emits \texttt{null} for uncertain fields, whereas the other agents more frequently mark capabilities as \texttt{ok}. Since $C_5$ counts false positives (claimed pass but probe fails), conservative abstention can reduce $C_5$ without improving runtime success. For Claude Code backends, most $C_5$ errors are capability-level rather than path/version mismatches, implying that failures are less about incorrect static facts and more about inferring readiness from installation logs without executing the corresponding probes.

\begin{table*}[!t]
    \centering
    \scriptsize
    \setlength{\tabcolsep}{3pt}
    \renewcommand{\arraystretch}{1.1}
    \begin{tabular}{lrrrrrrrrr}
        \toprule
        \textbf{Agent} &
        \textbf{\shortstack{$C_0$\\missing/total $\downarrow$\strut}} &
        \textbf{\shortstack{$C_1$\\CPU SR\strut}} &
        \textbf{\shortstack{$C_2$\\CUDA SR\strut}} &
        \textbf{\shortstack{$C_3$\\1-GPU SR\strut}} &
        \textbf{\shortstack{$C_4$\\DDP SR\strut}} &
        \textbf{\shortstack{$C_5$\\path\strut}} &
        \textbf{\shortstack{$C_5$\\version\strut}} &
        \textbf{\shortstack{$C_5$\\cap.\strut}} &
        \textbf{\shortstack{$C_5$\\total $\downarrow$\strut}} \\
        \midrule
        Codex (GPT-5.1-Codex) & 675/2858 (23.6\%) & 17/29 (58.6\%) & 35/44 (79.5\%) & 19/43 (44.2\%) & 11/32 (34.4\%) & 0 & 0 & 4 & 4 \\
        Claude Code Agent (GLM-4.7) & 761/2858 (26.6\%) & 16/29 (55.2\%) & 40/44 (90.9\%) & 21/43 (48.8\%) & 12/32 (37.5\%) & 0 & 1 & 17 & 18 \\
        Claude Code Agent (Sonnet 4.5) & 728/2858 (25.5\%) & 15/29 (51.7\%) & 41/44 (93.2\%) & 18/43 (41.9\%) & 12/32 (37.5\%) & 0 & 0 & 20 & 20 \\
        NexAU (DeepSeek-V3.1-Nex-N1) & 722/2858 (25.3\%) & 16/29 (55.2\%) & 41/44 (93.2\%) & 21/43 (48.8\%) & 11/32 (34.4\%) & 1 & 1 & 14 & 16 \\
        \bottomrule
    \end{tabular}
    \caption{Main results on ResearchEnvBench. $C_0$ reports missing imports over total imports after synthesis; $C_1$--$C_4$ report stage success rates (success/total on applicable repositories). $C_5$ counts hallucinations by type (Path/Version/Capability) and totals them ($\downarrow$ is better).}
    \label{tab:main_results}
\end{table*}

\Needspace{6\baselineskip}
\section{Analysis}

\subsection{Failure Mode Analysis: The Gap Between Existence and Readiness}
\label{sec:failure_modes}

We repeatedly observe an \emph{existence--readiness gap}: agents can often produce an environment that looks plausible under shallow checks (dependencies installed, imports largely resolve, GPUs visible), yet the repository's \emph{native} training/inference entrypoints still fail. The failures rarely come from missing a single ``core'' framework; instead, they concentrate in brittle, repository-specific assumptions---native extensions, auxiliary toolchains, and mixed-framework stacks---that are under-specified in manifests and only revealed on the real execution path.

\paragraph{The funnel of failure.} Tab~\ref{tab:main_results} also illustrates stage-wise success rates across $C_1$--$C_4$ for four agent settings. Even minimal CPU execution ($C_1$) succeeds on only 51.7--58.6\% of CPU-applicable repositories. While CUDA alignment ($C_2$) is relatively high (79.5--93.2\%), success drops at single-GPU execution ($C_3$, 41.9--48.8\%) and remains low for multi-GPU distributed execution ($C_4$, 34.4--37.5\%). Although stage denominators differ due to applicability, the consistent gap between ``GPU-visible'' checks and true entrypoint execution indicates that accelerator visibility is a necessary but insufficient condition for repository-level executability.

\paragraph{Anatomy of missing dependencies.} To understand why repository entrypoints still fail after ``reasonable'' installs, we categorize \texttt{ModuleNotFoundError} traces from failed runs. Missing modules follow a long-tail pattern and concentrate in a few recurring categories:
\begin{itemize}
    \item \textbf{Native extensions and accelerated operators.} Many repositories depend on CUDA/C++ operators that must be built (or matched to the correct PyTorch/CUDA application binary interface (ABI)), not merely \texttt{pip install}. Typical failures include \texttt{mmcv.\_ext} (\textit{facebookresearch/sapiens}), \texttt{flash\_attn} (\textit{GeeeekExplorer/nano-vllm}), and \texttt{lmcache.c\_ops} (\textit{LMCache/LMCache}).
    \item \textbf{Auxiliary tooling.} Utilities such as \texttt{wandb} and \texttt{tensorboard} can be hard dependencies on the default training path. Humans often fix these with a quick install; agents frequently break late and fail to recover.
    \item \textbf{Mixed-framework stacks.} Some repositories rely on multiple frameworks. When agents assume a single-framework world, they miss dependencies such as \texttt{jax} or \texttt{tensorflow}, causing cross-framework pipelines to fail.
\end{itemize}

Overall, agents tend to optimize for \emph{generic setup} rather than \emph{path-aware readiness}. The dominant blockers---native operators, auxiliary tooling, and mixed-framework dependencies---define the practical gap between ``it installs'' and ``it runs''.

\subsection{Capability Hallucination}
Agents are also unreliable reporters. We compare each agent's self-report (\texttt{report.json}) against hidden probes and observe a recurring ``overconfidence'' pattern (Table~\ref{tab:main_results}): agents often infer readiness from clean installation logs rather than verified execution.

\paragraph{Path \& version mismatches.} Even static facts can drift. We observe \emph{Path Hallucination}, where the reported Python executable does not exist (e.g., claiming \texttt{/usr/bin/python} when it is absent), and \emph{Version Hallucination}, where agents report assumed versions (e.g., \texttt{torch 2.1.0+cu121}) that differ from what the package manager actually installed.

\paragraph{False positives on runtime readiness.} The most damaging case is \emph{Capability Hallucination} ($C_5$): agents claim capabilities such as \texttt{cuda\_available=True} or \texttt{ddp\_expected\_ok=True} without running the corresponding smoke tests. In \textit{GeeeekExplorer/nano-vllm}, for example, the agent reported inference readiness, yet the first probe failed due to a missing \texttt{flash\_attn} kernel. This gap shows that ``installation succeeded'' is not a reliable proxy for ``the entrypoint runs''.

\subsection{Efficiency and Resource Cost}

Beyond success rates, agent benchmarks must account for efficiency. Table~\ref{tab:resource_consumption} reports per-repository averages for Codex (GPT-5.1-Codex), Claude Code (GLM-4.7), Claude Code (Sonnet 4.5), and NexAU (DeepSeek-V3.1-Nex-N1).

Two observations stand out. First, more tokens do not translate to commensurate readiness: NexAU consumes $\sim$20$\times$ more tokens (957.1k vs.\ 48.4k) yet matches Codex on $C_4$ success rate (both 34.4\%). Second, the average setup time and environment footprint differ only moderately across settings. For example, Claude Code (Sonnet 4.5) matches Claude Code (GLM-4.7) on $C_4$ while using more setup time (13.3 vs.\ 9.8 min) and a slightly larger environment (6.88 vs.\ 6.65 GB). While this comparison confounds model and agent design (not just budget), the lack of improvement despite a much larger interaction budget is consistent with a regime where remaining failures are dominated by a small set of build- and ABI-sensitive dependencies rather than incremental interaction alone.

\begin{table}[!t]
    \centering
    \scriptsize
    \setlength{\tabcolsep}{3pt}
    \renewcommand{\arraystretch}{1.15}
	    \begin{tabularx}{\linewidth}{@{}>{\raggedright\arraybackslash}X r r r r@{}}
	        \toprule
	        \textbf{Agent} & \textbf{\shortstack{Avg.\\Tokens (k)}} & \textbf{\shortstack{Avg. Setup\\Time (min)}} & \textbf{\shortstack{Avg. Env\\Size (GB)}} & \textbf{\shortstack{$C_4$ Success\\Rate (\%)}} \\
	        \midrule
	        Codex (GPT-5.1-Codex)  & 48.4 & 6.8 & 6.06 & 34.4 (11/32) \\
	        Claude Code (GLM-4.7) & -- & 9.8 & 6.65 & 37.5 (12/32) \\
	        Claude Code (Sonnet 4.5) & -- & 13.3 & 6.88 & 37.5 (12/32) \\
	        \mbox{NexAU (DeepSeek-V3.1-Nex-N1)} & 957.1 & 9.2 & 6.55 & 34.4 (11/32) \\
	        \bottomrule
	    \end{tabularx}
    \caption[Resource Consumption Comparison]{Resource Consumption Comparison. Avg. Tokens (k) is the total prompt+completion tokens consumed during synthesis. Avg. Setup Time is agent-phase wall time. Avg. Env Size comes from the env-size stage (\texttt{env\_\allowbreak size\_\allowbreak avg\_\allowbreak gb}). $C_4$ SR is computed as success/(total$-$skipped), excluding repos where $C_4$ is inapplicable.}
    \label{tab:resource_consumption}
\end{table}

Disk footprint is a proxy for how targeted an agent's dependency decisions are. Under uncertainty, agents (e.g., Claude Code backends) often install broadly to hedge. This rarely fixes a missing ABI-sensitive operator, and it can increase the chance of version skew and conflicts. Section~\ref{sec:case_sapiens} illustrates this failure mode.

\subsection{Case Study}
\label{sec:case_sapiens}

To ground the above discussion, we analyze \textit{facebookresearch/sapiens}, a representative repository where a plausible CUDA-visible stack still fails due to an implicit native-operator dependency.

\paragraph{Failure signature.} Both agents pass CUDA alignment ($C_2$), yet fail immediately when the training path imports \texttt{mmcv.ops} and cannot load the native extension \texttt{mmcv.\_ext} (\texttt{ModuleNotFoundError}). This blocks $C_1$ and consequently prevents $C_3$/$C_4$ execution.

\paragraph{Root cause.} \texttt{mmcv} may appear ``installed'' under shallow checks (e.g., \texttt{import mmcv} succeeds), but the real training path requires compiled CUDA/C++ operators. Satisfying this dependency requires selecting or building an ABI-matched variant for the target PyTorch/CUDA stack---an implicit build step that is rarely spelled out in manifests.

\paragraph{Cost.} Both agents spend substantial budget while remaining stuck on a single unfulfilled native dependency. GPT-5.1-Codex spends 12.4 minutes and produces an 11.67\,GB environment; NexAU spends 10.3 minutes and produces an 11.59\,GB environment. Yet both fail on the same missing operator, showing that ``installing more'' does not resolve build-bound dependencies.

\paragraph{Takeaway.} Sapiens highlights that generic ``install a plausible DL stack'' behavior is insufficient for research-grade repositories with hardware-bound operators. Robust environment synthesis requires path-aware smoke tests along the true import/entry path, and build-aware dependency resolution for native extensions; otherwise, agents can enter expensive installation loops that never address the actual blocking condition.

\section{Conclusion}
ResearchEnvBench targets a missing capability in current agent evaluations: synthesizing research-ready environments for modern AI/HPC repositories under hardware constraints. By enforcing a pyramid of runtime verification ($C_0$--$C_4$) and auditing self-reports ($C_5$), we reveal a recurring gap between ``GPU-visible'' stacks and repository-level executability, driven by implicit native operators and brittle version coupling. This benchmark will accelerate progress toward agents that can not only install dependencies, but also validate and report runtime readiness reliably.

In the future, we will extend ResearchEnvBench along three directions. First, we will broaden execution targets beyond our current standardized single-container CUDA sandbox to more realistic deployment settings, including repositories that ship official containers, multi-container stacks (e.g., Compose/Kubernetes), and cluster environments. This will let us evaluate environment synthesis that must reconcile in-container dependency resolution with host- and scheduler-level constraints rather than assuming a clean container-only bootstrap. Second, we will move from smoke-test verification toward more workload-faithful checks, including short training/inference runs under minimal synthetic inputs, asset-availability-aware probes (e.g., gated weights/datasets), and stability tests (e.g., determinism and numerical sanity) under explicit time and resource budgets. Third, we will strengthen the coupling between agent reports and executable evidence by requiring auditable artifacts (pinned manifests/lockfiles, command traces, and reproduction scripts) for \texttt{ok} claims, and explore calibration and learning signals that reduce $C_5$ capability hallucinations without sacrificing $C_1$--$C_4$ executability.



\bibliographystyle{unsrtnat}
\bibliography{references}

\end{document}